\documentclass[aps,prl,twocolumn,showpacs,superscriptaddress]{revtex4-1}

\usepackage{amssymb,amsmath}
\usepackage{graphicx}
\usepackage{dcolumn}
\usepackage{natbib}
\usepackage{longtable}
\usepackage{color}

\begin{document}

\title{Universal lineshape of the Kondo zero-bias anomaly in a quantum dot}

\author{Andrey V. Kretinin}
\email{andrey.kretinin@weizmann.ac.il}
\affiliation{Braun Center for Submicron Research, Condensed Matter Physics Department, Weizmann Institute of Science, Rehovot, Israel}
\author{Hadas Shtrikman}
\affiliation{Braun Center for Submicron Research, Condensed Matter Physics Department, Weizmann Institute of Science, Rehovot, Israel}
\author{Diana Mahalu}
\affiliation{Braun Center for Submicron Research, Condensed Matter Physics Department, Weizmann Institute of Science, Rehovot, Israel}

\date{\today}

\begin{abstract}
Encouraged by the recent real-time renormalization group results we carried out a detailed analysis of the nonequilibrium Kondo conductance observed in an InAs nanowire-based quantum dot and found them to be in excellent agreement. We show that in a wide range of bias the Kondo conductance zero-bias anomaly is scaled by the Kondo temperature to a universal lineshape predicted by the numerical study. The lineshape can be approximated by a phenomenological expression of a single argument $eV_{sd}/k_{\rm{B}}T_{\rm K}$. The knowledge of an analytical expression for the lineshape provides an alternative way for estimation of the Kondo temperature in a real experiment, with no need for time consuming temperature dependence measurements of the linear conductance.
\end{abstract}


\maketitle

In quantum dots an unpaired spin of the localized electron may act as a single magnetic impurity, which interacts with electrons in metallic leads forming a Kondo-correlated state \cite{Glazman1988,*Ng1988}. The quantum dots have been recognized as a versatile system for probing the many-body nature of the Kondo effect \cite{Goldhaber-Gordon1998a,*Cronenwett1998,Goldhaber-Gordon1998}. The main advantage of quantum dot hosting the Kondo state is the ability to tune its main characteristic parameter, the Kondo temperature ($T_{\rm{K}}$). Thereby, it is possible to study the reaction of the many-body system to various external perturbations, such as temperature $T$, bias $V$ and magnetic field.

One of the remarkable properties of the Kondo effect is that the response of the Kondo-enhanced conductance to a perturbation is governed by a set of universal laws independent of the physical system in which the Kondo state is realized. It has been experimentally verified that in the low-energy limit ($\{T, e|V|/k_{\rm{B}}\}<<T_{\rm K}$, where $k_{B}$ is the Boltzmann constant and $e$ is the elementary charge) the Kondo conductance as a function of $T$ and $V$ is described by a universal quadratic law \cite{Grobis2008,*Scott2009,*Yamauchi2011,Kretinin2011} expected from the Fermi-liquid theory \cite{Costi1994,Nozi'eres1974,*Schiller1995}. At higher temperatures ($T \sim T_{K}$) the Kondo conductance has been shown to scale to a universal dependence of a single parameter $T/T_{K}$ found from numerical renormalization group (NRG) study \cite{Costi1994}. These NRG calculations were later approximated by the phenomenological expression \cite{Goldhaber-Gordon1998} of a single parameter $T/T_{K}$, which became the main method for estimation of the Kondo temperature in real systems. Recently, we also attempted to check the universal behavior of the magnetic field dependence in both the low- and intermediate-field range \cite{Kretinin2011}. Despite all the theoretical advances, the problem of the nonequilibrium Kondo model at intermediate bias ($|V| \sim k_{\rm{B}}T_{\rm{K}}/e$) remained unsolved. As a result, the lineshape of one of the main hallmarks of the Kondo effect in quantum dots, known as zero-bias conductance anomaly (ZBA) \cite{Wingreen1994} had not been fully described by the theory. Only very recently, the 2-loop real-time renomalization group (RTRG) calculations developed for the Kondo model \cite{Pletyukhov2012} provided a numerical description of the nonequilibrium Kondo conductance in a wide range of biases. The result of these calculations agreed very well with the experimental data and linked previously developed analytical theories made in the low- \cite{Costi1994} and high-energy \cite{Haldane1978} limits. The development of the above mentioned RTRG calculations gave us an additional impulse to study the Kondo ZBA in more detail and test the universality of its lineshape.

In this Letter we present a detailed study of ZBA associated with the Kondo effect observed in an InAs nanowire-based quantum dot. The analysis of our experimental data shows that in a wide range of applied bias the Kondo ZBA can be scaled into a universal lineshape predicted by the RTRG calculations \cite{Pletyukhov2012}. We found our experiment to be well described by the proposed phenomenological expression, which accurately approximates the RTRG results in the experimentally relevant range of bias. We also show that this expression can be employed as an alternative method of quick and accurate estimation of the Kondo temperature.

\begin{figure}[!ht]
  \includegraphics[width=1\columnwidth]{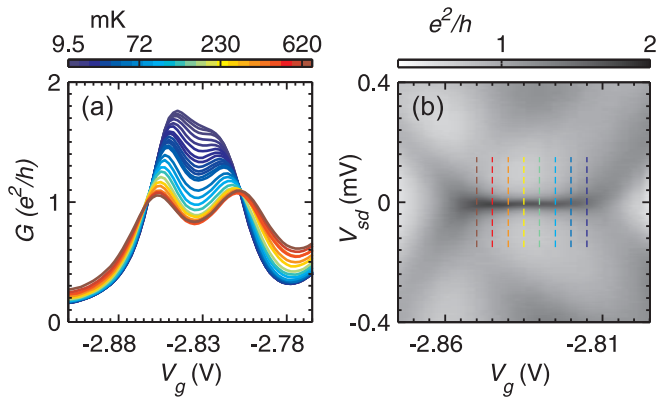}\\
  \caption{(a) The temperature dependence of the linear conductance observed in the InAs nanowire-based quantum dot. (b) The two-dimensional gray-scale plot of the nonequilibrium conductance measured in $V_{g}-V_{sd}$ plain at $T=T_{\rm{base}}$ for the same range of $V_{g}$ as in (a). The colored vertical dashed lines mark the conductance traces at fixed values of $V_{g}$ used for the scaling analysis ($V_{g} = -2.85$~V brown, $V_{g} = -2.845$~V red, $V_{g} = -2.84$~V orange, $V_{g} = -2.835$~V yellow, $V_{g} = -2.83$~V green, $V_{g} = -2.825$~V cyan, $V_{g} = -2.82$~V light blue, $V_{g} = -2.815$~V dark blue).}\label{Fig1}
\end{figure}
\emph{Experiment.}--The quantum dot used in the experiment was formed in a 50~nm-diameter high-quality InAs nanowire grown by the vapor-liquid-solid method on a (011) InAs substrate using molecular-beam epitaxy. The as-grown nanowires were randomly deposited onto a $p^{+}$-Si/SiO$_{2}$ substrate and individually connected by ohmic (Ni/Au) source and drain electrodes using e-beam lithography. To avoid undesirable effects of the underlying substrate the nanowires were suspended in vacuum over grooves predefined in the substrate, and fixed from the sides by the contacts. The lateral size of the dot was defined by the contacts separation ($\sim$450~nm) due to the electrons in nanowire being localized between two Schottky barriers formed at the nanowire-contact interface. The experiment was performed in a He$^{3}$-He$^{4}$ dilution refrigerator with the base temperature $T_{\rm{base}} \approx $10~mK. The transport measurements were made by a standard lock-in technique. Depending on the temperature the ac excitation bias was kept smaller or equal to $k_{B}T$. The dc bias $V_{sd}$ was applied to the source with respect to the drain, "virtually" grounded through the transimpedance preamplifier. The dot occupancy was tuned by the backgate voltage, $V_{g}$, applied to the $p^{+}$-Si substrate. More details on growth, fabrication and experimental set-up can be found in Ref.~\onlinecite{Kretinin2011}.

\begin{figure}[!t]
  \includegraphics[width=1\columnwidth]{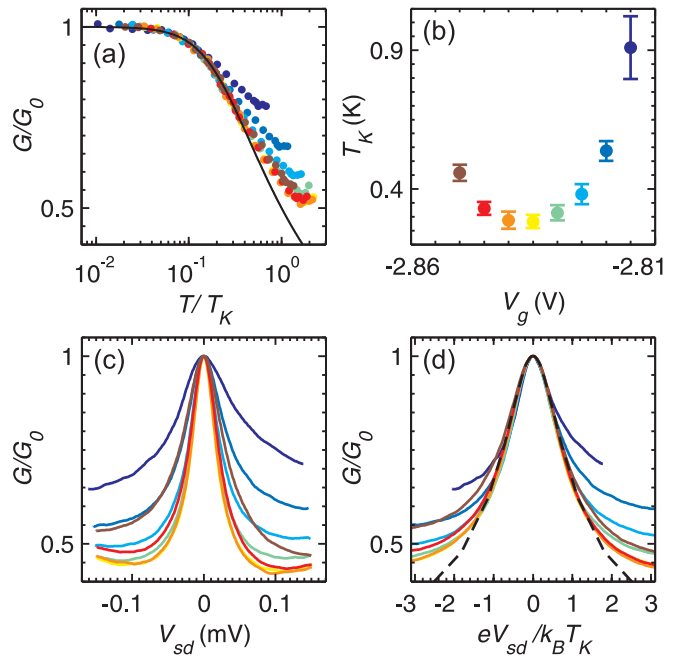}\\
  \caption{(a) Scaled temperature $G(T)$ dependence measured at different $V_{g}$ (colored circles) [see Fig.~\ref{Fig1}(b) for color code], approximated with Eq.~(\ref{DGG_formula}) (solid black curve). (b) The dependence of the estimated value of $T_{\rm{K}}$ on $V_{g}$. The error bars here represent the 68\% confidence interval. (c) The normalized conductance $G/G_{0}$ as a function of $V_{sd}$ measure at different $V_{g}$ and $T=T_{\rm{base}}$. (d) The same normalized conductance traces as in (c), but replotted as a function of $eV_{sd}/k_{B}T_{K}$ (colored curves). The traces are shown to collapse onto the same universal dependence given by RTRG calculations \cite{Pletyukhov2012} (black dashed curve). }\label{Fig2}
\end{figure}
\textit{Large bias scaling.}--To investigate the Kondo ZBA in more details we used the experimental results previously examined in our study of the low-bias and magnetic field scaling of the spin-1/2 Kondo conductance [see Ref.~\onlinecite{Kretinin2011}, Fig.1(c)]. Figure~\ref{Fig1}(a) presents the linear differential conductance $G$ through the quantum dot as a function of $V_{g}$ taken at different $T$. The region of $V_{g}$ at around $-2.83$~V corresponds to the conductance enhanced at lower temperatures due to the presence of the many-body Kondo state. There are two Coulomb blockade peaks emerging at higher temperatures, which mark the region of $V_{g}$ with an odd dot occupancy. To identify the relevant Kondo ZBA in Fig.~\ref{Fig1}(b) we show the gray-scale plot of the nonequilibrium differential conductance measured for the same range of $V_{g}$. The ZBA is seen here as a black horizontal line at around $V_{sd} =$~0. To test the theoretically predicted scaling of the Kondo ZBA we chose several nonequilibrium conductance traces taken at fixed values of $V_{g}$ as shown in Fig.~\ref{Fig1}(b) by colored dashed lines. For each of the chosen values of $V_{g}$ we estimated $T_{\rm{K}}$ by fitting the temperature dependence of the linear conductance $G(T)$ with the phenomenological expression \cite{Goldhaber-Gordon1998}
\begin{equation}
\textstyle
    G(T)/G_{0} = \left[1+\left(T/T'_{\rm{K}}\right)^{2}\right]^{-s},
    \label{DGG_formula}
\end{equation}
where $G_{0} = G(T = 0, V_{sd} = 0)$, $T'_{\rm{K}} = T_{\rm{K}}/(2^{1/s} - 1)^{1/2}$ and the parameter $s =$~0.22. Here the definition of $T_{\rm{K}}$ is such that $G(T=T_{\rm{K}},V_{sd}=0)=1/2G_{0}$. The experimental $G(T)$ dependencies fitted with Eq.~(\ref{DGG_formula}) with $T_{\rm{K}}$ used as a fitting parameter are shown in Fig.~\ref{Fig2}(a). The circles represent the experimental points taken at the values of $V_{g}$ color coded in Fig.~\ref{Fig1}(b), and the solid black curve corresponds to Eq.~(\ref{DGG_formula}). The deviation from the theoretical curve is related to additional mechanisms engaged in the transport at higher temperature. To avoid the effect of additional mechanisms the fitting procedure was made only for $T \leq$~200~mK. The fact that at $T <$~200~mK all the experimental data plotted as a function of $T/T_{\rm{K}}$ collapse on the same theoretical curve reflects the universality of the temperature dependence of the Kondo conductance \cite{Costi1994,Goldhaber-Gordon1998}. The estimated value of $T_{\rm{K}}$ at chosen $V_{g}$ is shown in Fig.~\ref{Fig2}(b) and follows the previously observed \cite{Wiel2000,Kretinin2011} parabolic-like dependence \cite{Haldane1978}.

As noticed in the early experiments \cite{Goldhaber-Gordon1998a,Wiel2000,Sasaki2000,*Nygard2000,*Jespersen2006,*Csonka2008} the width of the Kondo ZBA appears to be proportional to $T_{\rm{K}}$; the same qualitative behavior is observed in our experiment. Figure~\ref{Fig2}(c) shows the normalized nonequilibrium conductance $G/G_{0}$ at different $V_{g}$ with a well-pronounced ZBA maximum at $V_{sd} =$~0. The widest ZBA (dark blue curve) is associated with the highest $T_{\rm{K}} \approx$~900~mK, and the narrowest ZBA (orange and yellow curves) corresponds to the lowest $T_{\rm{K}} \approx$~300~mK [see Fig.~\ref{Fig2}(b)]. In accordance with the RTRG calculations \cite{Pletyukhov2012} the nonequilibrium conductance in the Kondo regime is scaled by $T_{\rm{K}}$ into a universal dependence (these calculations were made in the zero-temperature limit, as for our experiment, $T/T_{\rm{K}} \leq$~0.03, which was theoretically checked \cite{Pletyukhov2012} to be close enough to zero). To verify this prediction we plotted the normalized conductance $G/G_{0}$ as a function of the scaled bias $eV_{sd}/k_{\rm{B}}T_{\rm{K}}$ in Fig.~\ref{Fig2}(d). It is evident from Fig.~\ref{Fig2}(d) that all curves collapse onto the same universal dependence, which is given by the RTRG calculations and plotted as the dashed black curve. The deviations from the prediction observed at higher biases occur due to the approaching resonant level of the dot. In this case the system switches from the Kondo to the mixed-valence regime where the scaling is no longer valid \cite{Goldhaber-Gordon1998,Schoeller2000}.

\begin{figure}[!t]
  \includegraphics[width=1\columnwidth]{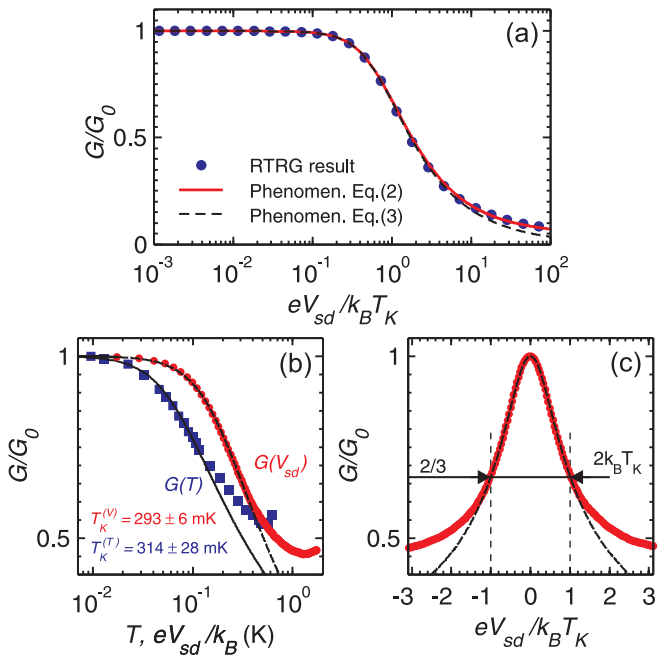}\\
  \caption{(a) The RTRG calculations of the nonequilibrium Kondo conductance as a function of bias taken from Ref.~\onlinecite{Pletyukhov2012} (blue circles) and its approximation with Eq.~(\ref{HS&MP_formula}) (red solid curve) and Eq.~(\ref{simple_formula}) (black dashed curve). (b) Extraction of $T_{K}$ using two different measurements, $G(T)$ (blue squares) and $G(V)$ (red circles) fitted with Eq.~(\ref{DGG_formula}) (black solid curve) and Eq.~(\ref{simple_formula}) (black dashed curve), correspondingly. The two extracted values $T^{(T)}_{\rm{K}}$ and $T^{(V)}_{\rm{K}}$ are the same within the statistical error. (c) The experimental Kondo ZBA measured at $V_{g}=$~-2.83~V [see green dashed line in Fig.~\ref{Fig1}(b)] (red circles) and its theoretical approximation made with Eq.~(\ref{simple_formula}) (black dashed curve). When the bias is such that $e|V_{sd}|=k_{\rm{B}}T_{\rm{K}}$ the Kondo conductance decreases to 2/3 of its zero-bias value $G_{0}$.}\label{Fig3}
\end{figure}
\textit{Phenomenological formula.}--As shown above the ZBA in the Kondo regime appears to have a universal dependence on bias, scaled by $T_{\rm{K}}$. This property of ZBA makes it potentially valuable for extracting $T_{\rm{K}}$ from the bias dependence of the conductance $G(V_{sd})$, in the same way as for $G(T)$ by approximating it with Eq.~(\ref{DGG_formula}). Here the use of numerical calculations is quite cumbersome, and the authors of Ref.~\onlinecite{Pletyukhov2012} suggested a phenomenological formula for describing their RTRG results, which we utilized for our definition of $T_{\rm{K}}$ \footnote{ The original formula in Ref.~\onlinecite{Pletyukhov2012} is derived for Kondo temperature $T^{*}_{\rm{K}}$ defined as $G(eV_{sd}=k_{\rm{B}}T^{*}_{\rm{K}})=1/2G_{0}$ and $(T^{*}_{\rm{K}}/T_{\rm{K}})^{2}\approx \pi$. One can obtain the formula for $T^{*}_{\rm{K}}$ by substituting $\pi$ with 1 in the denominator of Eq.~(\ref{HS&MP_formula}).}
\begin{equation}
    G(T=0, \nu)/G_{0} = \left[1+\frac{\left(2^{1/s_{1}}-1\right)\nu^{2}}{\pi+b\left(|\nu|^{s_{2}}-1\right)}\right]^{-s_{1}},
    \label{HS&MP_formula}
\end{equation}
where $\nu = eV_{sd}/k_{\rm{B}}T_{\rm{K}}$. The best fit of Eq.~(\ref{HS&MP_formula}) to the RTRG results gives $b = 0.05 \pm 0.01$, $s_{1} = 0.32 \pm 0.01$ and $s_{2} = 1.28 \pm 0.03$, Fig.~\ref{Fig3}(a) illustrates the quality of this fit. As seen from the plot, Eq.~(\ref{HS&MP_formula}) (red solid curve) approximates the numerical results (blue circles) very well in a wide range of bias. However, in real experimental situation the Kondo ZBA rarely extends beyond $\pm 2k_{\rm{B}}T_{\rm{K}}/e$ due to a dominating background, thus for the narrower range of bias ($e|V_{sd}|/k_{\rm{B}}T_{\rm{K}} < 10$) one can use a simplified expression
\begin{equation}
    G(T=0, \nu)/G_{0} = \left[1+\alpha \nu^{2}\right]^{-s_{1}},
    \label{simple_formula}
\end{equation}
where $\alpha = (2^{1/s_{1}}-1)/\pi$. The simplified formula is identical to the one in Ref.~\onlinecite{Pletyukhov2012}. As can be seen from Fig.~\ref{Fig3}(a) the simplified expression, shown by the black dashed curve, approximates the numerical result adequately up to $eV_{sd}/k_{B}T_{\rm{K}}\sim10$. Note that in the low-bias limit $e|V_{sd}|/k_{\rm{B}}T_{\rm{K}}\ll1$ Eq.~(\ref{simple_formula}) reduces to the quadratic Fermi-liquid dependence $G(V_{sd})/G_{0} \approx 1 - s_{1}\alpha(eV_{sd}/k_{\rm{B}}T_{\rm{K}})^2 = 1 - c_{V}(eV_{sd}/k_{\rm{B}}T_{\rm{K}})^2$, where the coefficient $c_{V}\approx0.78$ is in a good agreement with both theory and experiment \cite{Kretinin2011}.

Since the Kondo ZBA is scaled by $T_{\rm{K}}$ into a universal lineshape, which is described by a phenomenological formula, it should be possible to perform the reverse operation and extract the value of $T_{\rm{K}}$ by fitting the experimental $G(V_{sd})$ dependence with Eq.~(\ref{simple_formula}). Figure~\ref{Fig3}(b) shows a semilogarithmic plot of the Kondo conductance measured as a function of $T$ (blue circle) and the modified bias $eV_{sd}/k_{\rm{B}}$ (red circles). Each set of data is approximated with its own phenomenological formula using $T_{\rm{K}}$ as a fitting parameter; $G(T)$ with Eq.~(\ref{DGG_formula}) for $T\leq$~200~mK and $G(V_{sd})$ with Eq.~(\ref{simple_formula}) for $|V_{sd}|\leq$~30~$\mu$V, correspondingly \footnote{Approximation of $G(V_{sd})$ with the unsimplified Eq.~(\ref{HS&MP_formula}) gives the same value of $T_{K}$ as Eq.~(\ref{simple_formula}) within its statistical error.}. The two values of the Kondo temperature were found to be $T^{(T)}_{\rm{K}} = 314\pm28$~mK for $G(T)$ and $T^{(V)}_{\rm K} = 293\pm6$~mK for $G(V_{sd})$ and they are equal within the statistical error. Also, fitting of $G(V_{sd})$ provides statistically more accurate value of $T_{\rm K}$ due to a larger number of experimental points available for analysis. In our opinion this method of estimating $T_{K}$ may be advantageous to the traditional one involving measurement of the $G(T)$ dependence and approximating it with Eq.~(\ref{DGG_formula}). First, it is much easier to reliably measure the $G(V_{sd})$ dependence at the lowest possible temperature, rather than performing time-consuming measurements of $G(V_{g})$ in the linear regime at multiple temperatures. Normally, the low temperature ($T/T_{\rm K}<<1$), required for Eq.~(\ref{simple_formula}) is easily satisfied, especially at submillikelvin temperatures, since for most systems $T_{\rm K}\geq$~200~mK. Second, as pointed out, the value of $T_{K}$ extracted from the experimental $G(V_{sd})$ dependence is potentially more accurate.

\textit{Width of ZBA.}--Finally, we would like to discuss an important practical implication of Eq.~(\ref{simple_formula}). So far, the width of the Kondo ZBA has been used as a rough experimental estimate for $T_{\rm K}$ \cite{Goldhaber-Gordon1998a,*Cronenwett1998,Wiel2000,Sasaki2000,*Nygard2000,*Jespersen2006,*Csonka2008} and the full width at half maximum (FWHM) of the ZBA peak was assumed to be approximately equal to $2k_{\rm B}T_{\rm K}$. However, more careful measurements demonstrated an overestimation of $T_{\rm K}$ determined by this method, if compared to the values extracted from the $G(T)$ dependence \cite{Wiel2000}. To clarify the issue of the ZBA width we did a simple mathematical analysis of Eq.~(\ref{simple_formula}), which revealed that $G(V_{sd}=\pm k_{\rm B}T_{\rm K}/e) = 0.67G_{0} \approx 2/3G_{0}$ and the ZBA peak FWHM~$=2k_{\rm B}T^{*}_{\rm K} \approx 2\sqrt{\pi}k_{\rm B}T_{\rm K}$ \cite{Note1}. By means of these simple relations a quick and accurate estimation of $T_{\rm K}$ from the raw data is straightforward. As an example Fig.~\ref{Fig3}(c) shows the Kondo ZBA peak measured at $V_{g}=$~-2.83~V and plotted as a function of the normalized bias $eV_{sd}/k_{\rm B}T_{\rm K}$, where $T_{\rm K}$ is found from fitting of the same data with Eq.~(\ref{simple_formula}) [see Fig.~\ref{Fig3}(b)]. The universal lineshape is shown in Fig.~\ref{Fig3}(c) by the black dashed curve, illustrating an excellent agreement with the experiment. It is seen that the width of the Kondo ZBA is $2k_{\rm B}T_{\rm K}$ at two third of its total magnitude, contrary to the earlier assumed half magnitude \cite{Wiel2000,Nygard2000,*Jespersen2006,*Csonka2008}. It also suggests an explanation to the discrepancy between the FWHM and the value of $T_{\rm K}$ reported in Ref.~\onlinecite{Wiel2000}, alternative to the dephasing by bias \cite{Wingreen1994,Kaminski2000}. Unfortunately, we were unable to measure the FWHM reliably because in our experiment the Kondo ZBA at $0.5G_{0}$ was already broadened by the background.

In conclusion, using recent results of RTRG calculations of the nonequilibrium Kondo model we experimentally verified the universal scaling of the Kondo conductance at intermediate ($e|V_{sd}|\sim k_{\rm B}T_{\rm K}$) bias. We established that the Kondo ZBA in a quantum dot can be scaled by $T_{\rm K}$ to a universal dependence predicted by the numerical calculations and approximated by the phenomenological formula. An excellent agreement with the experiment allowed us to use this formula to extract the value of $T_{\rm K}$ solely from the analysis of the Kondo ZBA lineshape at the lowest temperature. This method appears to be quicker and more statistically accurate compared to the traditional one involving the measurement of the $G(T)$ dependence. Also, a closer look at the phenomenological formula revealed that when the applied bias is equivalent to $k_{\rm B}T_{\rm K}/e$ the Kondo conductance is at two thirds of its zero-bias value. At the same time the FWHM is about $2\sqrt{\pi}k_{\rm B}T_{\rm K}$, which is larger than it was previously thought to be. We demonstrated that those relations can provide an immediate and accurate estimate of the experimental $T_{\rm K}$.

The authors would like to thank Moty Heiblum for making this research possible and for suggestions and critical remarks. We acknowledge Herbert Schoeller and Mikhail Pletyukhov for providing the results of RTRG calculations prior to their publication and for critical reading of the early version of the manuscript. We also thank Yuval Oreg and David Goldhaber-Gordon for enlightening discussions. This work was partially supported by the EU FP6 Program Grant~506095, by the Israeli Science Foundation Grant~530-08 and Israeli Ministry of Science Grant~3-66799.

\bibliography{manuscript}

\end{document}